\documentclass[11pt,preprint]{aastex}

\usepackage{epsf}
\usepackage{rotating}

\def\spose#1{\hbox to 0pt{#1\hss}}

\def\lax{$\mathrel{\spose{\lower 3pt\hbox{$\mathchar"218$}}
     \raise 2.0pt\hbox{$\mathchar"13C$}}$}
\def\gax{$\mathrel{\spose{\lower 3pt\hbox{$\mathchar"218$}}
     \raise 2.0pt\hbox{$\mathchar"13E$}}$}

\def\lta{\mathrel{\spose{\lower 3pt\hbox{$\mathchar"218$}}
     \raise 2.0pt\hbox{$\mathchar"13C$}}}
\def\gta{\mathrel{\spose{\lower 3pt\hbox{$\mathchar"218$}}
     \raise 2.0pt\hbox{$\mathchar"13E$}}} 
 
\def\ergcm2s{${\rm erg~cm^{-2} s^{-1}}$}
\def\ergscm2s{${\rm erg~cm^{-2} s^{-1}}$}

\def\cm2{${\rm cm^{-2}}$}
 \def\ergs{${\rm erg~s^{-1}}$}

\def\sax1808{SAX~J1808.4-3658}
\def\msol   {{M$_{\odot}$}}
\def\lax    {${_<\atop^{\sim}}$}
\def\gax    {${_>\atop^{\sim}}$}

\begin{document}

\title{Resolved Jets and Long-Period Black Hole X-ray Novae}

\author{M. R. Garcia\altaffilmark{1}, 
J. M. Miller\altaffilmark{1}, 
J. E. McClintock\altaffilmark{1}, 
A. R. King\altaffilmark{2},
J. Orosz\altaffilmark{3}}

\affil{Harvard-Smithsonian Center for Astrophysics, 60 Garden Street, Cambridge, MA 02138}
\affil{Department Physics and Astronomy, University of Leicester, University Road, Leicester LE1 7RH, UK}
\affil{Department of Astronomy, San Diego State University, 5500 Campanile Drive, San Diego, CA 92182-1221}
\email{garcia,jmmiller,jem@cfa.harvard.edu, 
ark@astro.le.ac.uk,
orosz@zwartgat.sdsu.edu
}		

\begin{abstract}

In this brief note we point out that the four spatially resolved
relativistic jets among the 14 dynamically confirmed black hole
X-ray novae are all in systems with long orbital periods.  Many
shorter period systems show transient radio outbursts which are
attributed to jets, but these jets have not been spatially resolved.
Super-Eddington accretion has been suggested as a requirement for jet
formation and may be consistent with our compilation of luminosities, 
but some super-Eddington outbursts did not form spatially resolved 
jets.  We speculate that some as yet unknown process (or combination
of processes) favors formation of substantially larger jets in long
period systems.  Two short period systems show evidence for extent,
but have not been resolved into multiple components as the long period
systems have.

\end{abstract}

\section{Introduction}

The discovery of relativistic jets in SS433 (Margon et al.\ 1979, Margon
1984) brought the study of jets into the Galaxy from its previous
extragalactic home.  The speed of these $v=0.26c$ jets was
emphatically exceeded with the discovery of $v=0.92c$ jets in
GRS~1915+105 by Mirabel \& Rodriguez (1994).  This ``microquasar'' was
the first galactic source to show apparent velocities exceeding the
speed of light.  This discovery led to a search for superluminal jets
in other galactic X-ray binaries, and $\sim$ a dozen examples have now
been found (Mirabel \& Rodriguez 1999).

Many X-ray binaries have been observed with radio interferometers
(Hjellming \& Han 1995).  While most binaries are unresolved by these
observations, the similarities in both rapid variability and the
spectral indices between resolved and unresolved sources has led to
the belief that relativistic jets are  commonplace in
X-ray binaries (Mirabel \& Rodriguez 1999, Fender 2002).  While it is
clear that jets occur in X-ray binaries with both neutron star and
black hole primaries, there is a trend for the most relativistic jets
(ie, those with $v>0.9c$) to be associated with black holes (Mirabel
\& Rodriguez 1999, Livio 1997, but see Fender 2002 for exceptions to
this trend).

Soft X-ray transients (SXTs) are a subclass of X-ray binaries which,
rather than being persistently bright at X-ray wavelengths, erupt from
near invisibility to become among the brightest objects in the sky at
X-ray wavelengths. They then fade away in \lax~1~year.  It is thought
that most SXTs ($\gtrsim 70\%$) contain black hole primaries (e.g.\
Charles 1998).  In particular, there are 14 systems where dynamical
mass estimates indicate the presence of compact objects that are too
massive to be neutron stars (e.g.\ see the recent review by Orosz
2002).  We will refer to these systems as the ``dynamical confirmed''
black hole X-ray nova (BHXN).  
Radio observations, which have the potential to detect and perhaps
resolve relativistic jets, have been made of nearly all of these.
Fender \& Kuulkers (2001) present a thorough summary of these
observations, noting which have been resolved.  They include in
addition the related fast transients and neutron
star transients (their Table 1). By separating the sources into those
containing neutron stars and those suspected of containing black
holes, they show that the black hole systems are more radio loud
than those containing neutron stars. 

In what follows we concentrate on these 14 BHXN, because the dynamical
data gives us insight into the nature (size, evolutionary state, mass
transfer rate) of these systems which is less available for the other
systems.  We separate the systems into those having long and short
orbital periods.  Among these 14 systems, 6~have periods longer than a
day.  This is an important dividing line among the systems, as the
nature of the secondary changes at this period.  For periods less than
a day the secondaries are close to the main sequence and the mass
transfer is driven by gravitational radiation and/or magnetic braking,
while at longer periods the secondaries have begun to evolve off the
main sequence and the mass transfer is driven by the nuclear evolution
of the secondary (which leads to higher time averaged mass transfer
rates; Menou et al.\ 1999; King, Kolb, \& Burderi 1996).
These evolved secondaries include stars which are in the Hertzprung
gap (GRO J1655-40, Kolb, King \& Ritter 1997) and low mass giants (GRS
1915+105, Greiner et al. 2001), and they all share the common property of being able
to drive high mass transfer rates. 

We note that by restricting our sample to dynamically confirmed BHXN,
we may introduce an observational bias: the secondary must be bright
enough to allow moderate S/N phase resolved optical spectra to be
obtained, which will be difficult for systems at the greatest
distances or with the smallest secondaries (ie, the shortest periods).
However, 8 of the 14~systems in the sample are short period systems
and their distances extend out to nearly the same range as the long
period systems (eg, XTE J1859+226 with a period of 9.2~hours is at
11~kpc, see Table 1). The similar numbers and distances for long and
short period systems indicates that the search for resolved jets is
not biased by this effect.

\section{Data}

We have gathered observational data on dynamically confirmed BHXN from
the literature and summarize it in Table~1.  The list is ordered by
orbital period (column 2), with the shortest period first (XTE J1118+480).  The
object with the longest period is GRS 1915+105, recently found to have
a period of 33.5 days (Greiner, Cuby \& McCaughrean 2001).  This was
also the first BHXN system found to have resolved relativistic
jets (Mirabel \& Rodriguez 1994).

The peak outburst luminosities (column 3) are generally from Narayan,
Garcia \& McClintock 1997 (hereafter NGM97) or Garcia et al.\ 1998 and
references therein, but have been adjusted to approximate the
bolometric ($\sim$0.1-100~keV) luminosity, as this is likely of most
interest to the theory of jet formation.  As the NGM97 compilation
covered the 1-40~keV range, this adjustment increases the luminosity
by a factor of $\sim 2$ for sources which show a ``soft excess''
(often modeled by multi-temperature disk black body) in their X-ray
spectrum.  We note that this adjustment is uncertain, as the low
energy cutoff of the detectors (and the ISM) renders the flux below
$\sim 1$~keV unobservable for all but one (XTE~J1118+480) of the sources
in our sample.  The distances (column 4) are taken from Lee, Brown \&
Wijers (2002), and the dynamical data (columns 5 and 6) is taken from
Orosz 2002, unless noted as otherwise in the table footnotes. We have
taken $L_{Edd} = 1.1\times 10^{38} ({\rm M_x/M}_{\odot})$, appropriate
for cosmic abundances and a moderate (neutron star like) redshift
correction (Galloway et al.\ 2002).  We assume a mass in the center of
the range listed in column 5.

In column 7 we indicate if the system has been resolved into multiple
components (MC). Throughout the following text we refer to
resolved sources as those which have been separated into two (or
more) spatially separate components.  We note that this is
qualitatively different from determining that a source is extended,
ie, larger than the beam of the detector.
Upper limits to the size of the radio counterparts are not always
quoted in the literature.  In these cases,  we quote an upper
limit (column 8) corresponding to the array beam size as listed in the on-line
documentation for each telescope given the array configuration at the
time of observation. 
The delay (column 9) is the time between the X-ray discovery and the
first radio interferometric observations.  Note that this time is 
shorter than the time required for the jets to reach the extent listed
in the previous column. 

We note that these observations span a large range of times after the
outburst, sometimes beginning within days, other times within weeks of the
outburst.  The observations often extend for a year or more.  If large scale jets
fade to undetectable levels with a few days of an outburst they may be
 missed in some objects.  For example, the September 1999
outburst of V4641 had an extraordinarily fast rise and fall at both
X-ray and radio wavelengths. The resolved relativistic jet was visible
only for the two days following the outburst, and the longer lasting
(3 week) remnant might not have been associated with a jet if not for
the prompt observations that detected the jet (Hjellming et al.\ 2000).  However,
relativistic jets have also been detected long after the initial X-ray
discovery, for example GRS~1915+105 continues to show resolved jets
many years after discovery \cite{fender.1915.1999}; the resolved jets
in GRO~J1655-40 were found $\sim 3$~weeks after the discovery of the
X-ray source \cite{zhang.1655.iauc, tingay.1655, hjellming.1655}; and
those in XTE J1550-564 are visible several years after they were
generated \cite{corbel.1550.nature}.  Therefore delays of a few days
in the start of observations should not bias the sample against
finding resolved, relativistic scale jets in the short (or long)
period systems.

\subsection{Notes on Individual Objects}

\subsubsection{Short Period Systems}

XTE J1118+480: Approximately 1 month after the start of the outburst,
high resolution imaging with MERLIN and a 65x35 mas beam found the
source unresolved (Fender et al.\ 2001).  At approximately the same
time, and in addition 2~months later, observations with the VLBA found
the source unresolved at the $\sim 1$~mas level (Mirabel et al.\ 2001).
This provides the strongest upper limit to the size of any radio jet
in a short period BHXN.  Given the association of jets with outbursts
near the Eddington luminosity, it is relevant to note that the outburst of
XTE~J1118+480 was unusually faint as compared to the other short
period systems.  The system never left the 'low-hard' state to enter
the more luminous 'high-soft' state.  Given the the observed power-law
spectrum, the 1-160~keV luminosity from McClintock et al.\ (2001) is
within $\sim 5\%$ of that in our fiducial bolometric band of
0.1-100~keV.

GRO~J0422+32: This BHXN was observed for over a year at the VLA in
configurations ranging from D-array to A-array, and from 2 to 20~cm.
There were no reports that the system was resolved, and it was well
detected at 2~cm during A-array observations 3~months after the start
of the X-ray outburst (Shrader et al.\ 1994).  As the
HPBW  of this configuration is $0.14''$, we list this as the upper
limit to the size of any jet. However, we note that the initial
observation of the radio source $\sim 10$~days after the X-ray
outburst (Han \& Hjellming 1992) had a
HPBW of $\sim 12''$ (VLA in D-array, detection at 6~cm wavelength).

The most complete coverage of the X-ray light 
curve was provided in the hard X-rays (20-200 keV) by CGRO/BATSE.
Observations at $\sim 1/2$
the peak luminosity with TTM allowed coverage down to 2~keV, and did
not show the existence of an extra soft component.  The maximum X-ray
luminosity listed in Table~1 covers the 2-600~keV band and should be
close to the bolometric luminosity (Esin et al.\ 1998).

GRS 1009-45: The 1993 outburst of GRS 1009-45 was not observed in the
radio, and no post-outburst radio observations have been reported
(Brocksopp et al.\ 2002, Masetti, Bianchini \& Della Valle 1997).  It is
therefore not possible to search for radio jets in this source.

We use the same method and parameters 
as described in Barret et al.\ 2000 to estimate the
distance to GRS~1009-45, but we use a new (fainter) measurement of the
quiescent magnitude ($R=21.2 \pm 0.2$)  from Filippenko et al.\ 1999.  
The largest uncertainty in the distance is due the lack of an accurate
spectral type, and we find distances from 7.5~kpc to 12.8~kpc for M0
to K5 surface fluxes.  Assuming a K8 spectra type we find a nominal
distance of 9~kpc. 

The peak X-ray luminosity is uncertain due to the lack of continuous
coverage below 20~keV.  The 1-10~keV flux as reported by Tanaka et al
1993 during an ASCA observation $\sim 2$~months after the X-ray peak
corresponds to a luminosity of $\sim 2 \times 10^{38}$\ergs\/ at the
9~kpc distance estimated by Filippenko et al.\ 1999.  The 20-100~keV
X-ray light curve (Chen et al.\ 1997) shows that the hard X-ray flux was
down from the peak by a factor of 10 at the time of this ASCA
observation, and also shows that the peak X-ray luminosity was
$1.8\times 10^{38}$\ergs . Scaling up the ASCA flux by 10$\times$ in
order to account for the decay and adding in the observed 20-100 keV
flux, we approximate the peak luminosity from 1-100 keV as $\sim 2
\times 10^{39}$\ergs, but note that the observed peak flux is only
10\% of this.  We note that soft excesses are often observed near the
peak of XRN outbursts, so even though there is no data to confirm its
presence at the peak of the GRS~1009-45 outburst such a component seems
likely given its presence during the decay.  We therefore approximate
the bolometric luminosity in Table~1 as twice the 1-100~keV
luminosity.

A0620-00: The 1975 outburst of A0620-00 pre-dated the VLA, but it was
observed with the NRAO Green Bank Interferometer with a $\sim 2''
\times 10''$ beam (Owen et al.\ 1976) and the Jordell Bank MkII-III
Interferometer (Davis et al.\ 1975).  At the time, there were no reports
that the radio source was resolved.  A subsequent investigation of the
hand drawn plots from Davis et al.\ (1975), along with a new estimate of
the observational errors, may provide evidence that the outburst was
resolved at $\sim 3''$ size (Kuulkers et al.\ 1999).  

At its peak, the $\sim$2-100 X-ray spectrum of A0620-00 showed a soft
component and a hard tail (Ricketts et al.\ 1975).  The peak 1-18~keV X-ray
luminosity was $1.3 \times 10^{38}$\ergs\/ at 1~kpc (Elvis
et al.\ 1975). From Figure 2 of Ricketts et al.\ (1975) we estimate the
hard tail would contribute only $\sim 10$\% to this total.  However,
below 1~keV the soft spectrum could contribute an additional $\sim 1.6
\times 10^{38}$\ergs\/ to the total luminosity, and we therefore list in
Table~1 a peak bolometric luminosity of $4 \times 10^{38}$\ergs\/ at a
revised 1.1~kpc distance.

GS 2000+25: VLA observations 9 days after the start of the outburst
failed to detect this source, but 22 to 60 days later it was
detected as an unresolved source. The observations used the VLA C/D
array, and detected the source at 1.49 Ghz to 14.9 Ghz (Hjellming et
al 1988).  We estimate a size of $<1.2''$ from the HPBW of the VLA C
array at 2~cm.

The peak X-ray luminosity is taken from NGM97 (1-40~keV),
but has been adjusted up by a factor of two to account for the soft
component in the spectrum (Tsunemi et al.\ 1989, Terada et al.\ 2002).

XTE~J1859+226: As in the case of A0620-00, this BHXN may prove to
invalidate our suggestion that resolved relativistic jets are only
seen in long-period systems.  Observations of the 1999 outburst at the
VLA show evidence of a faint structure $\sim 0.7''$ in extent
approximately 20 days after the X-ray outburst began (Brocksopp et
al.\ 2002).  Further details on this faint structure will appear in a
forthcoming paper (Brocksopp et al.\ 2002).

The peak X-ray flux listed in Table~1 is the bolometric accretion disk
flux calculated by Hynes et al.\ (2002) on 18~Oct~1999, but including
the factor of $\sim 2$ additional flux in the observed power-law
component, corrected for the 11~kpc distance and scaled to the ASM
peak of 1.5~Crabs seen two days earlier.

GRS~1124-683 (Nova Muscae 1991): Observations with the ACTA detected
this BHXN during a phase of decaying radio emission 10~days after the
outburst, and also during the rise of a second radio flare 20~days
after the outburst (Ball et al.\ 1995).  The source was not reported to
be resolved, and from the HPBW at 2~cm for the 2.5~km maximum
baselines used we estimate an upper limit to the source size of $2''$.

The maximum (bolometric) X-ray luminosity listed in Table~1 is the 1-40 keV luminosity
listed in NGN97, adjusted for a distance of 5.1~kpc (Gelino, Harrison
\& McNamara 2001) 
and increased by a factor of 2 in order to include the observed soft
component (Ebisawa et al.\ 1994, Greiner et al.\ 1994) that we assume
extends below 1~keV.

H1705-250 (Nova Oph 1977): There were no radio observations of the
only recorded outburst of this source in 1977 (Brocksopp et al.\ 2002).  VLA
observations of the source in quiescence during 1986 failed to detect
it, yielding an upper limit of $\sim 5$~mJy at
6~cm (Nelson et al.\ 1988). 

The X-ray luminosity is the 1-40~keV luminosity from NGM97, but
corrected by a factor of two in order to account for the bolometric
luminosity likely present in the soft component of the observed X-ray
spectrum (Cooke et al.\ 1984).

\subsubsection{Long Period Systems}

4U~1543-47: This BHXN has been observed to have outbursts in 1971,
1983, 1992, and most recently in 2002.  The brightest of these
outbursts was that in 1983 (Chen et al.\ 1997). Garcia et al. (1998)
estimate that the peak luminosity of this outburst was $4 \times
10^{39}$ (1-20~keV, at 8~kpc). The very soft X-ray spectrum (van der
Woerd et al.\ 1989) would indicate a bolometric luminosity $\sim
2\times$ larger.  This, and the revised distance estimate of 7.5~kpc
(Orosz et al.\ 2002c) would indicate a peak luminosity for this outburst
of $\sim 7 \times 10^{39}$\ergs .  There were no radio observations of
the 1992 outburst (Brocksopp et al.\ 2002).  Observations during
quiescence in 1998 did not detect a radio remnant from any of the
previous outbursts (Hunstead \& Webb 2002).

The 2002 outburst also showed a very soft spectrum (Miller \&
Remillard 2002).  It reached a peak rate in the ASM of $\sim 320$ c/s
(1 Crab = 73 c/s).
Assuming the spectrum was similar to that measured by Tenma in 1983
(thermal bremsstrahlung with kT=1.6~keV, Kitamoto et al.\ 1984) this
corresponds to a bolometric luminosity of $3.2 \times 10^{39}$\ergs at
7.5~kpc.
Observations of this outburst with the MOST did detect a radio
transient (Hunstead \& Webb 2002).  The source was not reported to be
extended. The modest beam size of MOST therefore indicates a size
\lax~1$'$.  Table~1 lists these parameters for the 2002 outburst.

XTE~J1550-564: 
Resolved ($\sim 5^{\prime\prime}$) 
relativistic ($v\sim 2c$) radio jets were observed shortly after the first 
strong X-ray flare in September 1998 (Hannikainen et al.\
2001).  Emission from material associated with these jets was 
observed in the radio and X-rays roughly 4 years 
later at angular separations from the source of $\sim 20^{\prime\prime}$
(Corbel et al.\ 2002).  In addition, Corbel et al.\ (2001)
suggested that the radio spectrum and variability during the decay of
the 1998 outburst indicated the presence of
much smaller jets on a mas scale.

The X-ray flux of XTE J1550-564 reached a peak during a bright flare
on 1998 Sept 19.  The power law dominated the spectrum at this time,
so it is not necessary to correct the bolometric flux for an
additional soft component. Table~1 lists the 1-20~keV flux from Orosz
et al.\ (2002b), which is within 20\% of that computed by Sobczak et
al.\ (2002) for the bolometric disk plus 2-100~keV power law
component.  We note that there is a strong soft component during the
plateau immediately following the bright flare, but the bolometric
flux at this time is substantially lower (Geirlinski \& Done 2002).

GRO J1655-40: Radio emission was not detected until $\sim 10$~days
following the 27~July~1994 X-ray outburst of this source (Harmon et
al.\ 1995, Zhang et al.\ 1994).  The radio source was not resolved until
$\sim 3$~weeks after the X-ray outburst (Tingay et al.\ 1995).  Nearly
3~months after the outburst it was visible at the VLA (1.3 cm) as a
pair of jets separated by up to $\sim 5''$ (Hjellming and Rupen 1995).
At 20-100~keV energies, this outburst appeared as five (or more)
discrete outbursts over more than 1~year. The first three of these has
simultaneous radio flares, but the last two were very radio
quiet (Rupen 2002).  The 1997 outburst also produced a radio flare,
but the VLA was in its most compact array (D) and was unable to resolve
structures on the size seen in 1994 (Rupen 2002).

GRO~J1655-40 is unusual among the long period systems in not reaching
an outburst luminosity of \gax~10$^{39}$\ergs .  The 1994/5 outburst
pre-dated the RXTE/ASM, so the entire outburst light curve was
accessible only at energies above 20~keV as afforded by CGRO/BATSE.
ASCA observations revealed the presence of a soft component in the
X-ray spectrum containing $\sim$40\% of the luminosity at a time when
the bolometric luminosity was estimated to be $\sim 10^{38}$\ergs\/
(Zhang et al.\ 1997, Zhang PC 2002).  The rates in the BATSE band were
2-5 times below their peak during this ASCA observation (Zhang et al
1997).  We therefore estimate that the bolometric luminosity at peak
was $5 \times 10^{38}$\ergs\, and list this value in Table~1.

V4641 Sgr: The maximum luminosity during the very fast, 12~Crab X-ray
flare lasting \lax~1~hour was $6\times 10^{39}$\ergs (2-100~keV), or
$3 \times 10^{39}$\ergs (2-10~keV) (Orosz et al.\ 2001).  This flare was
observed only with the RXTE/ASM, so it is not possible to determine if
there is an additional soft component which might increase the
bolometric luminosity beyond that in the 2-100~keV band.  At much lower
flux levels (0.01 and 0.3 Crab) I'nt Zand et al.\ (1999) note the presence of
an additional soft component, but Smith, Levine \& Morgan (1999) note
that the spectrum during the 12~Crab flare is ``harder than the Crab''.
We therefore assume  the 2-100~keV flux is close to bolometric, and list
this in Table~1.

V4641 Sgr was observed with the VLA $\sim 16$~hours following the
bright 12~Crab flare, and was resolved into a jet with $0.25''$
extent.  Observations 2  days later showed that the jet had faded
below the detection limit, but the core (presumed point of origin of
the jets) was marginally resolved and remained visible for another
26~days (Hjellming et al.\ 2000).

GS 2023+338 (V404 Cyg): This very bright radio transient ($\sim 1$~Jy)
 was not resolved at the VLA despite extensive observations using many
 array configurations (Han \& Hjellming 1992).  Because the source
 was not resolved, we take as an upper limit to its size the HPBW of
 the A-array at 2~cm which is $\sim 0.14''$.  Rapid ($\sim$5 minute)
 fluctuations of the radio flux and the necessity to avoid the Compton
 catastrophe require a source size of $\sim 0.2$mas (Han \& Hjellming
 1992).

The outburst spectrum of GS~2023+338 was unusual in that it did not
show an obvious soft thermal component (Tanaka 1989), instead being
consistent with a power-law with photon index $\sim 1.7$ (Terada et al
1994).  Careful review of the GINGA spectrum shows that the 
$\sim 1$~keV soft thermal component did make a brief appearance, perhaps at
sub-peak luminosities, but was shrouded by heavy absorption and/or
comptonization (Zycki, Done \& Smith 1999).

The flux appeared to reach a maximum saturation level (perhaps equal
to the Eddington limit) early on 30~May 1989, at which time the soft
component showed a temperature of $\sim 0.2$~keV.  The bolometric
luminosity 
at this time is computed to be
$\sim 3 \times 10^{39}$\ergs\/ (Zycki, Done \& Smith 1999) and this
value is listed in Table~1.  We note that the that the 1.7-37~keV
luminosity computed by Terada et al.\ (1994) assuming a simple power-law
model is $\sim 50$\% below our adopted value.

GRS 1915+105: The first reported detection of the radio counterpart of
this now prototypical microquasar was approximately 4~months after the
X-ray outburst began (Mirabel et al.\ 1992).  A large radio flare began
$\sim 1.5$~years after the discovery of the X-ray transient (Rodriguez
\& Mirabel 1993), which was subsequently resolved into relativistic
jets (Mirabel \& Rodriguez 1994).  

The maximum X-ray luminosity during bright (10s) flares
reaches an unabsorbed value of $7 \times 10^{39}$\ergs (1-200~keV).
The bolometric luminosity (if the emission extends down to 0.1~keV)
 may be as high as  $6 \times 10^{40}$\ergs (0.1-200~keV, Greiner
et al.\ 1998), and we list this value in Table~1.
This source continues to be active
at X-ray and radio wavelengths and continues to generate relativistic
jets (Fender et al.\ 1999). 

\clearpage
\newpage
\thispagestyle{empty}

\footnotesize 
\begin{sidewaystable}
\centerline{Table~1: 	Multiple Component Radio Jets in BHXN}
\hspace*{-0.9in}\begin{tabular}{lllllllccll} 
\hline

ID(outburst year) & Spectral & {${\rm P_{orb}}$} & {${\rm L_{max}/L_{Edd} }$} & D & ${\rm
 M_X}$ & $i$ & MC? & Size &Radio Obs &Delay\\
 &Type & hr  & erg~s$^{-1}$    & kpc & {\msol} & &  &  &    & \\\hline\hline
\multicolumn{11}{c}{Short Period Systems}\\\hline

XTE J1118+480(00) &K7V-M0V  & 4.1 & 36.1(1)/38.9 & 1.9   & 6.5-7.2     & $81 \pm 2$ & N & $<0.001''$(3)& VLA, VLBA, & 1~month(2,3,4) \\
	&	 &  &  &    &      &  &  & 						& Merlin,Ryle &  \\

GRO J0422+32(92)  & M0V & 5.1	& 37.7(45)/38.7 & 2.6(4) & 3.7-5.0    & $44\pm~2$   & N & $<0.14''$(5) 	&VLA	& 10~days(6)\\

GRS 1009-45(93)	& K6-M0 & 6.8  	& 39.6?(7,8)/38.6	& 9(9,25)  & 3.6-4.7?   &67? & N/A& N/A         &NONE	& N/A(10,11)\\

A0620-00(75)	& K4V & 7.8	& 38.6(12,13)/39.1 	&1.1(14)	& 8.7-12.9 &$40.8\pm 3$ & N(?) & \lax $3''$	& JBI,GBI &14~days(15,16)\\

GS 2000+25(88)	& K5V & 8.3	& 38.6(14,17)/38.9 &2.7  & 7.1-7.8   &$64.0\pm~1.3$ & N &  $<1.2''$ 	& VLA	&9~days(18)\\

XTE J1859+226(99) &... &9.2  & 38.9(19)/39.0 &11(20)   & 7.6-12.0?    &high & N(?)& \lax$ 1''$	& VLA, Merlin &2~days(11)\\

GRS 1124-683(91)  & K4V &10.4& 38.9(14,21)/38.9 &5.1(46)    & 6.5-8.2&	$54\pm 2$ & N &$<2''$	& ACTA, MOST &9~days(22)\\

H1705-250(77)	& K3V & 12.5& 38.6(14,23)/38.9  & 8.6  &  5.6-8.3     &	$>60$ & N/A& N/A 	&VLA	&11~years(24)\\\hline


\multicolumn{11}{c}{Long  Period Systems} \\\hline

4U 1543-47(02)& A2V & 26.8& 39.5(25)/39.0& 7.5(26)  & 8.4-10.4     & $20.7\pm~1.5$ & 	N &$<1'$	&MOST	&1.5day(27)\\

XTE J1550-564(98)& G8IV-K4III& 37.2& 39.2(28,29)/39.0 &	5.3  & 8.4-10.8&  $72\pm 5$ & 	Y & $>20''$& VLA, ACTA   & 2~days(30,31)\\
& &	&     &	     &  	    & 	  & & &MOST, VLBI,& \\
& &	&     &	     &  	    & 	  & & 	   &LBA& \\

GRO J1655-40(94)& F6III &62.4& 38.7(32)/38.8&	3.2  & 6.0-6.6      & $70.2\pm~1.2$ & Y & $0.1''$-$5''$ &VLA, VLBA & 10~days(33,34,35)\\

SAX J1819.3-2525(99)& B9III&   67.6&39.8(36)/38.9&	10   & 6.8-7.4 & $75\pm~2$ & Y & $0.25''$	&VLA, Merlin &1~day(37)\\
(V4641~Sgr) &  & & & & & & & \\

GS 2023+338(89) &K0IV & 155	& 39.4(38,39)/39.1 &3(40)    & 10.0-13.4    &   $56\pm~4$ & N  & $\sim 0.2$mas& VLA &  9~days(41)\\
(V404 Cyg)&  & & & & & & & \\

GRS 1915+105(92)& K-MIII& 804	& 39.8(42)/39.2 &12.5 & 10.0-18.0    & $70\pm~2$ &  Y & $>1''$	&VLA, Merlin &4~months(43,44)\\


\hline
\hline

\end{tabular} \\
\hspace*{-0.9in}\begin{minipage}{10.3in} 
Notes for Table:  The dynamical data (periods, masses, inclinations)
are taken from Orosz (2002). Distances and spectral types are
from Lee, Brown \& Wijers (2002), unless
otherwise noted.  Maximum outburst luminosities are from Garcia et al
1998 or NGM97 unless otherwise noted, but have been 
adjusted to be bolometric. The X-ray energy bands listed below correspond to the range
covered by the observations. Radio data is from a variety of sources,
often multiple. The 'Delay' refers to time between X-ray discovery and
first radio interferometric observations. Additional refs: 
(1)1-160~keV, McClintock et al.\ 2001,
(2)Fender et al.\  2001,
(3)Mirabel et al.\ 2001,
(4)Pooley \& Waldram 2000;
(5)Shrader et al.\ 1994,
(6)Han \& Hjellming 1992;
(7)Chen et al.\ 1997,
(8)Tanaka et al.\ 1993,
(9)Filippenko et al.\ 1999,
(10)Masetti et al.\ 1997,
(11)Brocksopp et al.\ 2002;
(12)2-18~keV, Elvis et al.\ 1975,
(13)2-80~keV, Ricketts et al.\ 1975,
(14)NGM97,
(15)Davis et al.\ 1975,
(16)Kuulkers et al.\ 1999;
(17)Terada et al.\ 2002,
(18)Hjellming et al.\ 1998;
(19)Hynes et al.\ 2002,
(20)Zurita et al.\ 2002;
(21)Greiner et al.\ 1994,
(22)Ball et al.\ 1995;
(23)Cooke et al.\ 1984,
(24)Nelson et al.\ 1988;
(25)This work,
(26)Orosz 2002c,
(27)Hunstead and Webb 2002;
(28)Orosz et al.\ 2002b,
(29)Sobczak et al.\ 2000,
(30)Hannikainen et al.\ 2001,
(31)Corbel et al.\ 2002;
(32)Zhang et al.\ 1997, Zhang P.C. 2002,
(33)Harmon et al.\ 1995,
(34)Tingay et al.\ 1995,
(35)Hjellming \& Rupen 1995;
(36)Orosz et al.\ 2001,
(37)Hjellming et al.\ 2000;
(38)Tanaka 1989,
(39)Zycki, Done \& Smith 1999, 
(40)Shahbaz et al.\ 1994,
(41)Han \& Hjellming 1992b;
(42)Greiner et al.\ 1998,
(43)Mirabel et al.\ 1992,
(44)Mirabel \& Rodriguez 1994
(45)2-600~keV, Esin et al.\ 1998,
(46) Gelino, Harrison \& McNamara 2001 
\end{minipage}
\end{sidewaystable}


\clearpage
\newpage

\normalsize

\section{Discussion}

From Table 1 we see that the four systems with well resolved (multiple
component) relativistic jets are XTE J1550-564, GRO J1655-40,
V4641~Sgr, and GRS~1915+105, which all have orbital periods
\gax~1~day.  The data also suggests that the radio counterparts of two
short period systems (A0620-00 and XTE~J1859+226) may be larger than
the beam size.  However, the data is unequivocal only in the case of these four
long period systems.  That is, relativistic jets have so far only been resolved
into two separate, moving components in long period systems.  It
appears that a long orbital period and bright outbursts are
necessary, but not sufficient, to give rise to resolved jets.  For
example, the long period systems V404~Cyg and 4U~1543-47 both have had
super-Eddington outbursts, but resolved jets have not been found
in either.

Given the small number of sources in the sample, we should ask what
the odds are that this apparent trend is merely a coincidence.  In
doing so, we should leave GRS~1009-45 (not observed) and H1705-250 (not
detected) out of consideration, as there are no observational 
constraints on them.  This leaves equal numbers of long and short
period systems.  Depending upon exactly how one asks the question, the
probability is $= 0.5^4 = 6.3\%$ (assuming each detection of a
resolved jet is equivalent to a trial with 50\% probability of
selecting a long or short period system) or $(6~ {\rm pick} ~4)/(12~ {\rm
pick} ~4) =
3\%$ (from the combinatorics of selecting 4 jets out of the 6 long
period systems vs. the number of ways to select 4 jets out of all 12
systems).  In either case the significance of the correlation is
suggestive ($3\% \sim 1.8\sigma$) but certainly not overwhelming.

Could observational biases lead to this correlation?  Certainly the
distances of the two classes of sources are not systematically biased,
so any observed differences in jet angular scales would indicate
physical size differences.  The resolved jets in the long-period
systems are typically \gax~$1''$ (V4641 Sgr, GRO J1655-40, GRS
1915+105) or even more (\gax~$20''$ in the case of XTE
J1550-564). Most of the upper limits to jets sizes in the short-period
systems are of a similar scale, ie, \lax a  few arcsec (GS~2000+25,
A0620-00, XTE~J1859+226, GRS~1124-683), but the limits for
GRO~J0422+32 and XTE~J1118+480 are factors of $\sim 10$ and $\sim
1000$ smaller.  However, the peak luminosities of these two systems
were $\sim 0.1$ and $\sim 0.001$ of Eddington, so neither had the
very bright outburst apparently required to produce resolved jets.

Radio jets can be associated with the transition from a quiescent to
outbursting state and may fade in a few days, (ie, V4641 Sgr,
Hjellming et al.\ 2000) in which case  prompt observations are required to
detect  them.  The delay between the X-ray outburst and the first
interferometric radio observations (column 10) ranges from 1~(2)~days
to 4~months (11~years) with a median of 5~(10)~days for the long
(short) period systems.  However, jets may also be visible for years
after the outburst has faded (ie, XTE J1550-564, Corbel et al.\ 2002)
indicating that prompt observations are not always required.
Brocksopp et al.\ (2002) tabulate the radio observations of 13 of the 14
sources in our sample, and note that the initial rise was observed in
the radio in only two systems, one with a short period (XTE J1859+226)
and one with a long period (XTE J1550-564).

There are several possible physical reasons to expect and/or explain
the correlation which may be present.  We discuss some of these
possibilities below, but conclude that the situation is currently unclear.

Could this difference be related to the smaller physical size
of the short period binary orbits?  Certainly the jets in AGN are
larger than those in microquasars. 
Assuming similar black hole masses,
GRO~J0422+32 and XTE~J1118+480 (the two shortest period systems, also
with the tightest limits to jet size) have orbital separations $\sim
5$x smaller than V4641~Sgr and GRO~J1655-40 (which have periods near
the median for the long-period sample).  If the longer period systems
have higher masses (Lee, Brown \& Wijers 2002) this difference would be
further enhanced, but even if we make the extreme assumption that they
have twice the mass of the short-period systems the orbital scales are
only $\sim 6$ times larger.  This difference is not negligible and
might explain an order of magnitude difference in the apparent jet
sizes.  Orbital dimensions alone cannot account for the
factor of \gax~1000 difference in the size of the jets in the long
period systems when compared to the best limit among the short period
systems (that in XTE J1118+480, of $<0.001''$), but given the very low
outburst luminosity of XTE~J1118+480 this comparison may not be
appropriate.

King (1998) suggested that jets appear in systems with 'Hertzprung
gap' secondaries, leading to super-Eddington mass transfer.  Others
(ie, Mirabel \& Rodriguez 1999, Fender 2001) have also suggested
that the presence of jets is related to super-Eddington mass transfer.
The data in Table~1 provides some support for this, in that the peak X-ray
luminosities of 3 of the 4 resolved sources are super-Eddington.  
The fourth (GRO J1655-40) is apparently not super-Eddington, although
it did reach within $\sim 70\%$ of its Eddington luminosity. However, given
the lack of continuous coverage below 20~keV one must allow the
possibility that this system did reach (unobserved) super-Eddington
luminosities. 
While the data allows that it may be necessary to reach
super-Eddington luminosities in order to produce resolved jets, it
does not appear to be sufficient.  The long period systems
V404~Cyg (39.5), 4U1543-47 (39.5) both were super-Eddington and did not
display resolved jets. The short period systems GRS~1124-683 (38.9) and
XTE~J1859+226 (38.9) both reached, but apparently did not exceed, the
Eddington luminosity.

Under low luminosity conditions, the inner edges of BHXN accretion
disks may be truncated long before they reach the last stable orbit by
evaporation of the disk into an advection or convection dominated flow
(Narayan et al.\ 1996, Hameury et al.\ 1997, Liu et al.\ 1999, Meyer-Hofmeister \& Meyer 2001).
It has been suggested that the rapid filling of this cavity in the
inner regions of the accretion disk is necessary to generate
relativistic jets (Mirabel \& Rodriguez 1999).  There is some
evidence that the size of this cavity may scale with mass, and that the
mass of the BH in the long period BHXN is larger than that in the
short period systems (Bailyn et al.\ 1998, Orosz et al.\ 2002b, Lee, Brown
\& Wijers 2002).  However, V4641 Sgr shows a strong Fe-line in
quiescence which may show the signature of relativistic broadening in
the inner disk, indicating that this BHXN does not have a large cavity
in the accretion disk during the low state (Miller et al.\ 2002).

Could the spins be higher in the long period systems, and could this
cause larger jets?  Blandford \& Znajek (1977) demonstrated that it is
possible, at least in principle, to extract rotational energy from a
black hole with high angular momentum through magnetic field lines
threading the horizon.  Observationally, magnetic connections to gas
in the plunging region (the region inside of which there are no stable
orbits) may be quite similar.  This may provide an energy source for
powering jets from black holes (see, e.g., Blandford 2001).  This has
some observational appeal: VLBI observations of M87 constrain the
point of jet collimation in this AGN to be within 100 Schwarzschild
radii  (Junor, Biretta
\& Livio 1999); this suggests a fairly compact region is involved in
forming the jet.  Moreover, this source is well-described by ADAF
models (e.g., Di Matteo et al. 2003), so the jet is not likely driven
by a disk in this source.

Evidence for significant angular momentum (or, black hole spin) has
also been implied in many Galactic black holes and black hole
candidates, but it is not yet clear if the spin has been accurately
measured in any source. This evidence has been found through a variety
of methods: measurements of the inner disk radius using continuum
X-ray spectroscopy (Zhang, Cui \& Chen 1997), via measurements of rapid
QPOs (Remillard et al.\ 2002), and via measurements of
relativistically broadened Fe emission lines (Miller et al.\ 2002).

The first method used to estimate the spin in a substantial number of
BHXN was to measure the inner radius of the disk via fitting the
continuum spectra to multi-color disk black body (MCBD) models
(Mitsuda et al. 1984).  Using this method Zhang, Cui \& Chen (1997)
found nearly maximal spins ($a\sim 1$) for GRO~J1655-40 and
GRS~1915+105 and near zero spins ($a\sim 0)$ for GS~1124-68 and
GS~2000+25.  
Hynes et al. (2002) derive an inner disk radius from
MCBD fits to XTE data that is consistent with a low spin for
XTE~J1859+226.

The MCBD method has some uncertainties that are not shared by the other
two methods.  Not only does the measured inner disk radius depend on the
distance to the source and the inclination of the inner disk (which
may be different from the system inclination), but the model itself
may fail to account for changing opacities due to Comptonization
effects (see Shimura \& Takahara 1995; see also Merloni, Fabian, \&
Ross 2000).  Regardless of the uncertainties in the theory, the
varying dominance of different spectral components may make it
difficult to do the spectral deconvolutions necessary to measure the
inner disk temperature and radius.  For example, while Zhang et al.\ (1997)
find a nearly maximal spin ($a = 0.93$) based on ASCA spectroscopy of
GRO~J1655-40, Sobczak et al.\ (1999) apply the same method to
a much larger RXTE dataset on the same source and find a much lower
spin ($a=0.5$) and a lower limit consistent with zero spin.
 
As the Keplerian orbital frequency at the marginally stable circular
orbit around a non-spinning black hole scales roughly as $\nu \simeq
220 {\rm Hz} (10~M_{\odot}/M_{BH})$, the observation of QPOs at
similar frequencies in systems with mass constraints may also provide
a means of measuring black hole spin.  Evidence for spin has been
revealed via this means in GRO J1655$-$40 (Strohmayer 2001,
$0.15<a<0.5$), in XTE J1550$-$564 (Remillard et al. 2002, $0.1<a<0.6$;
Miller et al. 2001 $a>0$), GRS~1915$+$105 (Morgan et al. 1997,
Strohmayer 2001b, Remillard et al.\ 2002, $a>0$), and perhaps
XTE~J1650$-$500 (Homan et al. 2002).

However, uncertainty in exactly what causes the QPO (ie, Homan et al
2001, Remillard et al.\ 2002) leads to uncertainty in the derived spin,
and values as high at $a>0.9$ may be allowed for XTE J1550-564
(Remillard et al.\ 2002) and GRO J1655-40 (Strohmayer 2001, Remillard et
al 2002).  With multiple QPOs in many systems, comparative measures of
the spin among different systems should likely be done by comparing
the systems when they are in similar luminosity and spectral states
(Mendez, Belloni \& van der Klis 1998).

Broad Fe~K$_{\alpha}$ emission lines -- sometimes asymmetrically
skewed and plausibly explained as emission lines produced at the inner
accretion disk -- can also provide evidence of spin.  At present, this
technique has produced strong evidence of spin only in XTE J1650-500
(Miller et al. 2002, $a\sim 1$).  While we do not include
XTE~J1650-500 in Table~1 because its binary parameters are not yet
well known,  this BHXN has a short orbital period of 0.212~days
(Sanchez-Fernandez et al.\ 2002).  A measurement of the Fe-line in
V4641~Sgr at a low flux level ($\sim 10^{-3} L_{Edd}$) finds a line
consistent with zero spin, but which does allow moderate spin (Miller
et al.\ 2002b).

At least one of each of the three methods discussed above has been
applied to each of the four jet systems in order to estimate its spin.
While the estimates do not always agree, there is evidence for
moderate to high spin in three of the four jet systems.  The spin in
V4641~Sgr may be somewhat lower.  Do the two long period systems
without resolved jets (V404 Cyg and 4U 1543-47) have even lower spin?
V404~Cyg did not show a strong soft disk component at the peak of its
outburst (Tanaka 1989), and while they did not fit MCBD models to the
spectra, Zhang, Cui \& Chen (1997) suggest that this lack of a disk
component may indicate a retrograde spin.  However, reanalysis of the
Ginga spectrum shows brief periods when there was a strong soft
component hidden beneath strong low energy absorption (Zycki et al
1999). The spectrum during each of the four recorded outbursts of
4U~1543-47 has contained a strong soft excess (van der Woerd 1989),
but thus far multi-color disk blackbody models have not been fit to
the spectrum.

Conversely, do the short period systems (where there are no resolved
jets) show evidence for lack of spin? All of the systems except for
XTE~J1118+480 and GRO~J0422+32 show a soft component in outburst,
indicative of an accretion disk possibly amenable to multi-color disk
black model fits.  The most interesting systems to explore would be
GS~2000+25, XTE~J1859+226, and GRS~1124-683, as they reached at least
$\sim 1/2$ of their Eddington luminosities and had $\sim 1''$
resolution radio imaging within about a week of their outbursts.
Hynes et al.\ (2002) found an inner disk radius of $\sim 90$km $\sim 6
{\rm R_g} $ for the 10\msol\/ black hole in XTE~J1859+226. While the
spectra did not allow tight constraints on this radius, it is
indicative of zero spin.  Zhang, Cui \& Chen (1997) have previously
noted that the X-ray spectra of GRS~1124-683 and GS~2000+25 are
indicative of near zero spin.  However, XTE~J1650-500 ($P_{orb} =
0.212$~days) has a highly skewed Fe line indicating a very high spin
($a\sim 1$, Miller et al.\ 2002).  Thus it is unclear if the short
period systems have uniformly lower spin than the long period systems.

As evidence for spin is often found in the systems with resolved jets,
it may be productive to explore whether the fastest jets observed may
be due to the highest spin parameters (Fabian \& Miller 2002).  At
present, robust measures of black hole spin are difficult to make.
Further observations of black holes with high time resolution and high
spectral resolution instruments may be able to make more robust
measurements.  It is possible that magnetic connections between the
disk and either the black hole or matter in the plunging region are
favored at high mass accretion rates in Galactic black holes; if this
is the case, spinning black holes may act to drive jets more often in
the more luminous long-period binaries.

\section{Summary}

In order to focus discussion, we have deliberately excluded several 
resolved relativistic jets in galactic sources which are not
dynamically confirmed BHXN.  In addition, it may not be fair to
compare the jets even amongst this smaller sample, as they may not all
be due to the same phenomena: the $>20''$ X-ray and radio jets in
J1550-564 may be powered by deceleration, while the rapidly evolving
jets from V4641 Sgr must be powered by relativistic electrons.
However, the data we present do suggest that long period BHXN produce
larger jets than short period systems, an observation which may
provide insight into the jet formation mechanism.  The data in Table~1
are consistent with the suggestion that an outburst luminosity \gax
Eddington is necessary but not sufficient to produce resolved,
relativistic jets.  Measurements of black hole spin are not yet
sufficiently robust to determine the role it plays in the formation of
resolved jets.  As has been often noted (ie, Mirabel \& Rodriguez
1999), the human time-scale of the evolution of micro-QSO jets may
allow insights that are difficult to gain from the much more slowly
varying extra-galactic jets.

\acknowledgments

We thank the Aspen Center for Physics for its hospitality during the
summer of 2002 which allowed work on this paper to begin.  MRG
acknowledges the support of NASA/LTSA grant NAGW-10889 and NASA
contract NAS8-39073.  JMM
acknowledges support from the NSF through its Astronomy and
Astrophysics Postdoctoral Fellowship Program. JEM acknowledges the
support of NASA ADP grant NAG5-10813. 
ARK gratefully acknowledges a Royal Society Wolfson Research Merit
Award.

\end{document}